# The GeoLifeCLEF 2023 Dataset to evaluate plant species distribution models at high spatial resolution across Europe

Christophe Botella [1], Benjamin Deneu [1], Diego Marcos [1,2], Maximilien Servajean [3], Théo Larcher [1], Joaquim Estopinan [1], César Leblanc [1], Pierre Bonnet [4], Alexis Joly [1]

1: INRIA, LIRMM, Montpellier

2: INRIA, Université de Montpellier, Montpellier

3: AMIS, Université Paul Valéry, LIRMM, Montpellier

4: CIRAD, UMR AMAP, F-34398 Montpellier, France

**Abstract:** The difficulty to measure or predict species community composition at fine spatio-temporal resolution and over large spatial scales severely hampers our ability to understand species assemblages and take appropriate conservation measures. Despite the progress in species distribution modeling (SDM) over the past decades, SDM have just begun to integrate high resolution remote sensing data and their predictions are still entailed by many biases due to heterogeneity of the available biodiversity observations, most often opportunistic presence only data. We designed a European scale dataset covering around ten thousand plant species to calibrate and evaluate SDM predictions of species composition in space and time at high spatial resolution (~ten meters), and their spatial transferability. For model training, we extracted and harmonized five million heterogeneous presence-only records from selected GBIF datasets and 6 thousand exhaustive presence-absence surveys both sampled during 2017-2021. We associated species observations to diverse environmental rasters classically used in SDMs, as well as to 10 m resolution RGB and Near-Infra-Red satellite images and 20 years-time series of climatic variables and satellite point values. The evaluation dataset is based on 22 thousand standardized presence-absence surveys separated from the training set with a spatial block hold out procedure. The GeoLifeCLEF 2023 dataset is open access and the first benchmark for researchers aiming to improve the prediction of plant species composition at a very fine spatial grain and at continental scale. It is a space to explore new ways of combining massive and diverse species observations and environmental information at various scales. Innovative AI-based approaches, in particular, should be among the most interesting methods to experiment with on the GeoLifeCLEF 2023 dataset.

# Introduction

Global changes transform ecosystems at an alarming rate (Diaz et al., 2019), but their local impact is highly context-dependent and difficult to predict. Regularly monitoring species composition at high spatial resolution (e.g. tens of meters) and continental or global scale is key to characterize the real time response of species assemblages and biodiversity indicators (e.g. diversity, intactness, habitat condition, presence of endangered species), but it has remained largely unfeasible. Species composition predictions across space and time from deep learning-based species distribution models (deep SDMs, Botella et al., 2018, Deneu et al., 2021) are a promising alternative as these models can efficiently exploit complex and high spatial resolution geographic predictors, including remote sensing data, to fill the sampling gaps (Deneu et al., 2022, Estopinan et al., 2022). However, the heterogeneity, imbalance and complexity of the available species observations and environmental data are major impediments to the implementation of species distribution models at this resolution.

Gathering standardized biodiversity data through presence-absence surveys in small plots is limited in spatial coverage and requires frequent costly updates. However, a promising alternative approach has emerged in the form of new biodiversity monitoring schemes, such as crowdsourcing programs (e.g. Pl@ntNet, iNaturalist, Observation.org). These programs yield millions of presence-only (PO) species records annually with precise geolocation, offering complementary insights. Despite their advantages and their growing use for species distribution models, PO records have certain limitations. They do not provide information about the absence of non-observed species (Hastie et Fithian, 2013). In addition, they only represent a fraction of the species communities in regions with limited sampling and are biased towards specific species (Feldman et al., 2021). As a result, incorporating these records into species distribution models can introduce numerous biases (see e.g., Boakes et al., 2010, Mesaglio et Callaghan, 2021). Therefore, while crowdsourced PO data opens up exciting possibilities, it is crucial to be mindful of these limits when using them for ecological analyses. In particular, the evaluation of species distribution models on PO data induces important evaluation biases only due to the sampling patterns, as pointed out by the previous GeoLifeCLEF campaign (Lorieul et al., 2022). This highlights the need for an evaluation procedure based on exhaustive sampling of species communities at high spatial resolution. In addition, a small number of standardized species observations, e.g. presence-absence (PA) plots, can also help solve many sampling biases of the PO data when jointly integrated into species distribution model calibration, while enabling to capitalize on the rich information hidden in the mass of PO data (Fithian et al., 2015, Miller et al., 2019). Even when using comprehensive PA data, it is difficult to model and map biological groups with large taxonomic diversity such as plants, including around 400 thousand species to date, a vast majority of which are rare.

Species distribution models have relied for decades on geographic predictors at a spatial resolution of the kilometer to a hundreds of meters, such as bioclimatic (Booth et al., 2014), land cover (Luoto et al., 2005) or human footprint variables (Di Marco et al., 2014). Besides, remote sensing data represent an unprecedented opportunity to provide high spatial resolution species distribution models with rich and globally-consistent predictor variables describing the environment (He et al, 2015) and its temporal changes, but its integration into species distribution models is recent and challenging (Deneu et al., 2022, Estopinan et al., 2022). This data is precious to complete the picture of the environmental landscape provided by the aforementioned variables at a coarser spatial scale, but integrating variables at different spatial or temporal resolutions within deep learning architectures brings its own set of challenges.

Open benchmark datasets are still rare in the domain of species distribution modeling, as is the objective comparison of methods on well established datasets (but see e.g. Valavi et al., 2022). Open benchmark datasets are becoming crucial with the increasing complexity of data available for modern species distribution models (Mouquet et al., 2015, Farley et al. 2018). Indeed, this complexity implies technical difficulties preventing their use, and a myriad of possible predictive methods to assess. To facilitate the use of the available diverse data and enable all researchers in ecological modeling and machine learning to focus on building innovative and useful predictive species distribution algorithms, we have assembled an European scale open dataset called GeoLifeCLEF 2023. It associates five millions of opportunistic PO records and thousands of standardized PA surveys of around ten thousand species. Species observations are aligned with diverse environmental data, including most variables classically used in species distribution modeling as well as high spatial resolution multi-band remote sensing images and time series, creating a suited playground to develop innovative deep learning models (**Figure 1**). The dataset importantly allows a fair and open evaluation of the predictive algorithms with a large test set of standardized presence-absence surveys. The evaluation scheme was designed in the context of the eponym model evaluation campaign hosted on Kaggle, to which seven international teams submitted 130 algorithms. Botella et al. (2023) explained the machine learning challenges of the campaign and compared the algorithm performances. Therefore, the dataset offers the possibility to directly compare new approaches with a diversity of state-of-the-art and documented approaches on an objective basis. Further models developed from this dataset will directly meet the needs of the Europe biodiversity strategy for 2030 through the HORIZON Europe projects MAMBO (10.3030/101060639) and GUARDEN (10.3030/101060693). They will be used in particular as a basis to map biodiversity indicators at European scale (e.g. presence of endangered species, invasive species, habitat condition metrics).

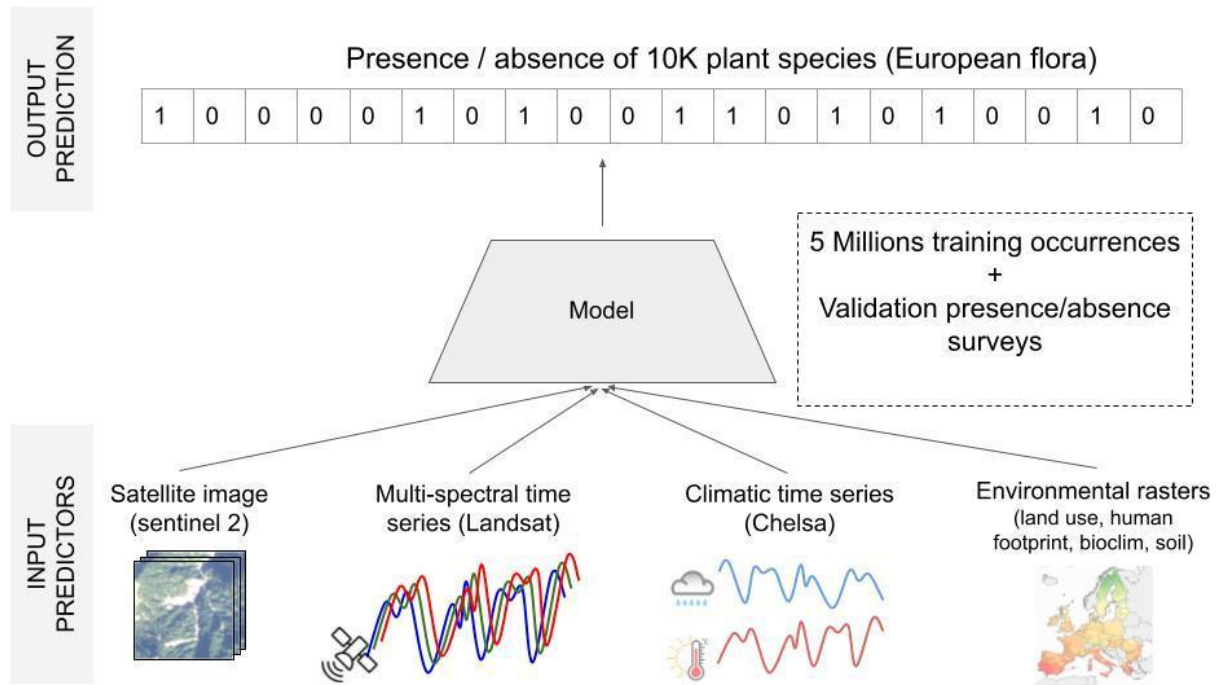

**Figure 1.** The GeoLifeCLEF 2023 dataset aims at developing and evaluating algorithms that predict the presence or absence of vascular plant species given a location and date based on environmental and remote sensing predictors. The dataset covers 38 European countries and more than ten thousand species.

# Dataset presentation

1. Study area

The full species observation data covers 38 European countries, namely the 27 European Union members, the U.K., Switzerland, Norway, the western balkan countries (Albania, Kosovo, Serbia, Bosnia & Herzegovinia, Montenegro, North Macedonia), Andorra and Liechtenstein. It doesn't include Greenland. Hence, this data covers 8 biogeographic regions (as defined by the European Environment Agency): Alpine, Atlantic, Black Sea, Boreal, Continental, Mediterranean, Pannonian and Steppic.

2. General data structure

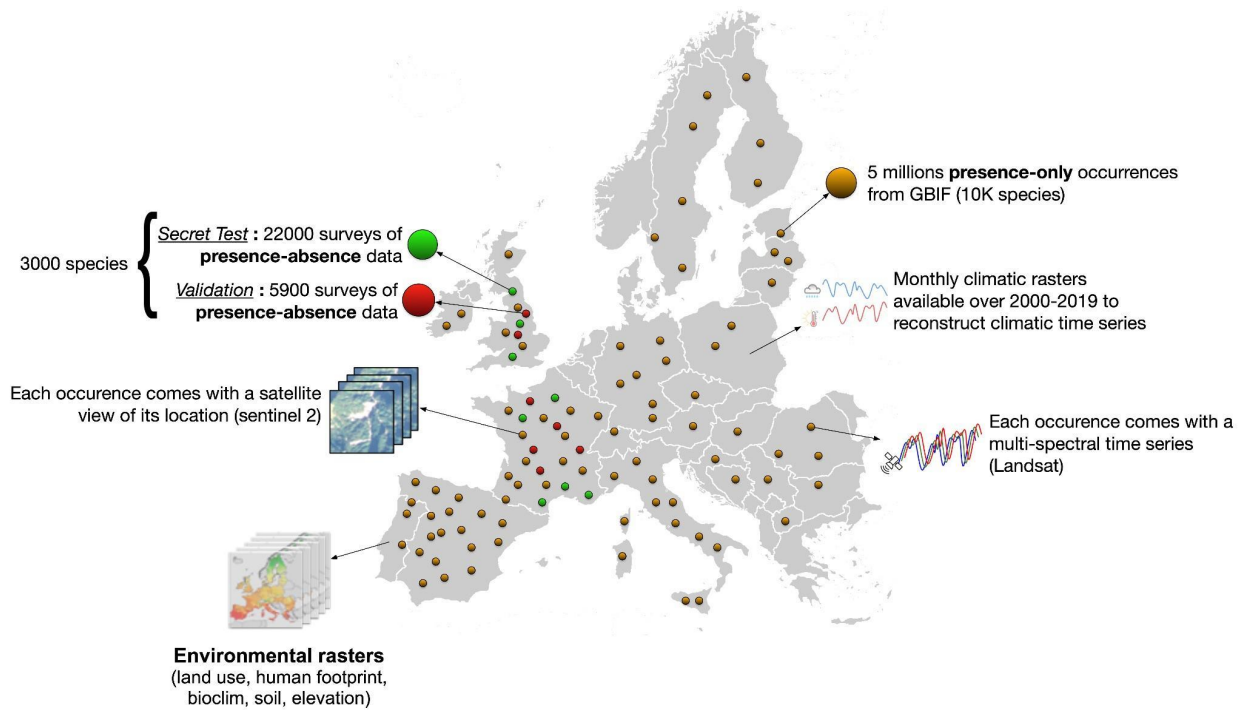

**Figure 2.** Presence only (PO) and presence-absence (PA) plant observation data are provided to train Species Distribution Models (SDMs) in order to predict plant species community composition at high spatial resolution from various environmental predictors (Satellite images, satellite time series, climatic time series and various environmental variables) with up to ten meter spatial resolution.

3. Species observation data

The training species observation data is composed on one hand of around five million plant species PO occurrences (species name, geolocation, time, dataset, observer, etc.) collected between 2017 and 2021 in Europe (38 countries), and on the other hand of around six thousand exhaustive presence-absence surveys (**Figure 2**). This data covers overall around ten thousand species, i.e. most of the European flora. This data should be used to inform the model outputs keeping in mind that the PO only partially represents the local species composition and it is subject to various sampling biases. The presence-absence data (PA) is provided to compensate for biases of PO data and in model calibration.

**Presence-only data.** The Presence-Only (PO) data are a set of 5 million records (one species identifier, geolocation, time etc.) and all extracted from the Global Biodiversity Information Facility (GBIF, extraction: ). This PO data originates from 13 pre-selected trusted source datasets (see **Table 1**) including three international citizen science programs (Pl@ntNet, iNaturalist RG,

Observation.org) and regional datasets, to favor a large spatial coverage across Europe (38 countries), as visible in **Figure 3**. The dataset pre-selection criteria are described in **Methods** below. All records were collected between 2017 and 2021 and had a geolocation uncertainty inferior to 100m, as reported on the GBIF. The PO data covers per se the 10,038 plant species observed in the full dataset, which include all species reported in the PA data. Despite the richness of this data, one should keep in mind that the local absence of a species among PO records doesn't mean it is truly absent: An observer might not have reported it because it was difficult to detect or identify at the given time, not a monitoring target or just not attractive. Indeed, most PO data come from citizen science programs without sampling protocol. This data is stored as a tabular CSV file named "Presences_only_train.csv" and its structure is described in section "**v - Dataset structure, file entries**" below.

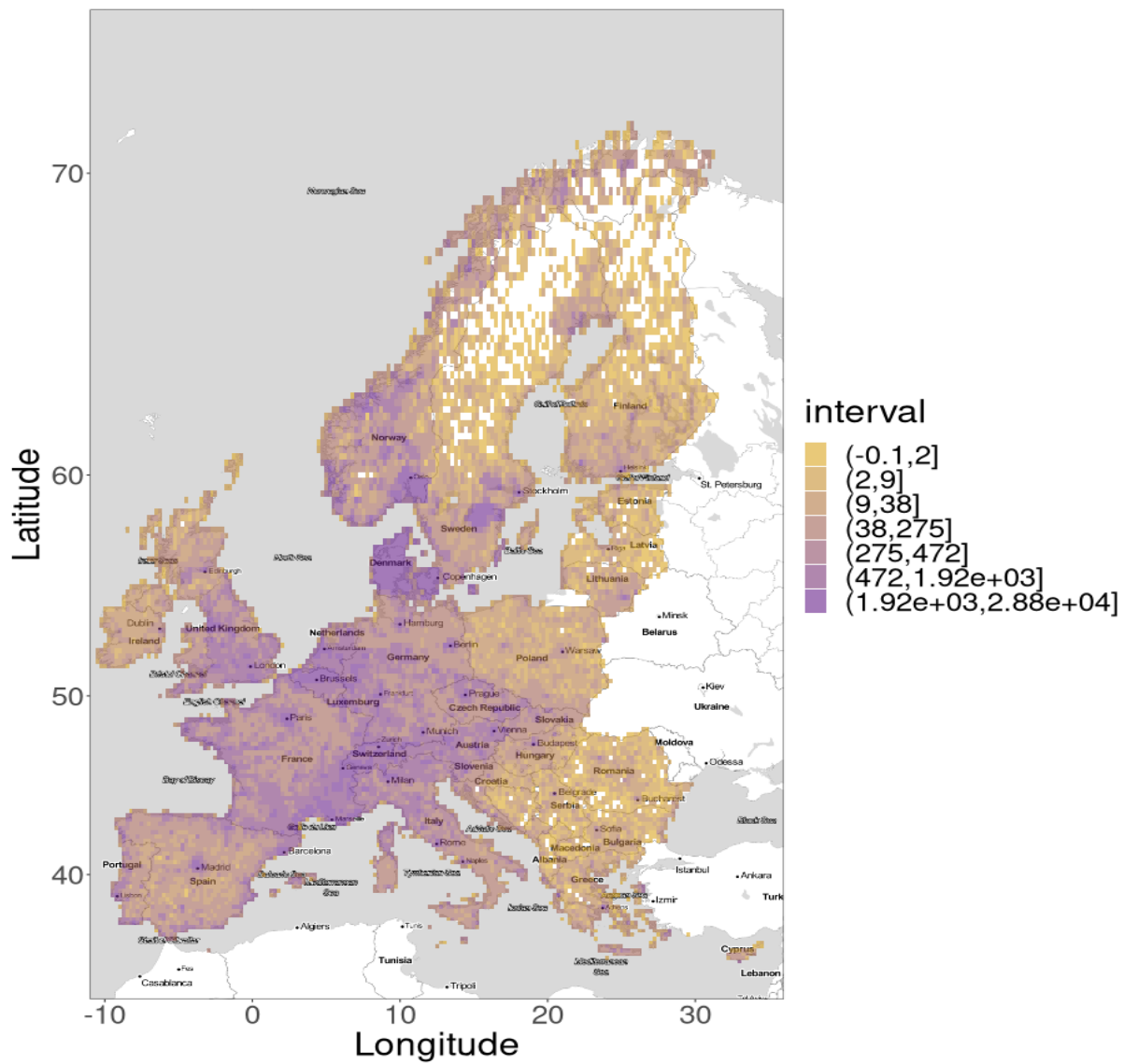

**Figure 3.** Count map of the five million presence-only observations at quarter degree scale. This data covers most of the European flora with 10,038 observed plant species.

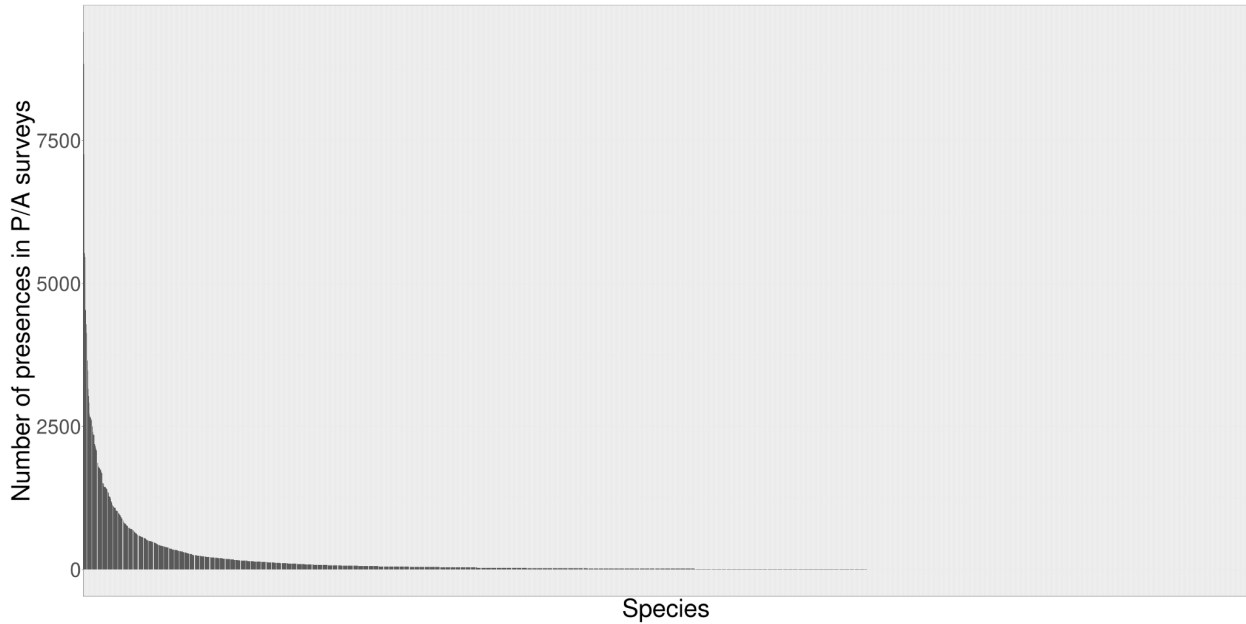

**Figure 4.** Number of presences in the Presence-Absence surveys (PA) per species by decreasing order of abundance: Most plant species are very rare. The graph shows the number of presences among 28 thousand PA surveys for the 3 thousand species observed in these surveys.

| GBIF Dataset name | Type | Source |
|---|---|---|
| Pl@ntNet automatically identified occurrences | Presence-only | GBIF |
| Pl@ntNet observations | Presence-only | GBIF |
| Swiss National Databank of Vascular Plants | Presence-only | GBIF |
| The Danish Environmental Portal, species and habitats-database "Danmarks Miljøportals Naturdatabase" | Presence-only | GBIF |
| iNaturalist Research-grade Observations | Presence-only | GBIF |
| Vascular plant records verified via iRecord | Presence-only | GBIF |
| Artportalen (Swedish Species Observation System) | Presence-only | GBIF |
| Waarnemingen.be - Non-native plant occurrences in Flanders and the Brussels Capital Region, Belgium | Presence-only | GBIF |
| Observation.org, Nature data from around the World | Presence-only | GBIF |

| Norwegian Species Observation Service | Presence-only | GBIF |
|---|---|---|
| Masaryk University - Herbarium BRNU | Presence-only | GBIF |
| Invazivke - Invasive Alien Species in Slovenia | Presence-only | GBIF |
| Données de l'inventaire forestier national de l'IGN - Relevés floristiques du protocole de l'Inventaire National Forestier de l'IGN | Presence Absence | GBIF |
| National plant monitoring scheme (England, Wales, Scotland) | Presence Absence | GBIF |
| Conservatoire Botanique National Alpin | Presence Absence | CBN-A |
| Conservatoire Botanique National Mediterranéen (SIMETHIS-Flore-CBNMed, Argagnon et al., 2022) | Presence Absence | CBNMed |

**Table 1.** The 16 source species observation datasets selected for GeoLifeCLEF 2023 include 12 Presence-Only (PO) datasets and 4 Presence-Absence (PA) datasets.

**Presence-Absence data.** The full Presence-Absence (PA) data is a set of 28K surveys (list of present species identifiers, geolocation, time, etc.). Each survey is an inventory of all the plant species that botanical experts could detect and identify on a given date with an important sampling effort over a small spatial plot of 10 to 400m². Thus, PA surveys report the local species assemblages with almost no observation bias. The PA records associated with one survey (longitude, latitude and day) implicitly inform on the absence of all non-observed species. As for the PO data, all PA surveys were sampled between 2017 and 2021 in France and Great Britain, with a geolocation uncertainty inferior to 100m. This data comes from four source datasets with different spatial extents or targeted habitats (see **Table 1** and **Methods**). Despite the large size of this PA dataset, it only covers 3043 species, i.e. approximately half of the western Europe flora, and these species observations are extremely imbalanced with most of them being only observed once or twice among all the PA surveys (see **Figure 4**). Therefore, the PO data may help to compensate for the lack of observations of many species, while the PA surveys may help to control for sampling biases in model calibration (Fithian et al., 2015, Miller et al. 2019). We used a spatial block hold-out procedure (Roberts et al., 2017) to randomly split the PA surveys between a validation set of 5,948 surveys (open access for model calibration) and a test set of 22,404 surveys (kept secret for fair evaluation) based on a spatial grid of 50x50km cells (see details in **Methods**) resulting in the spatial distribution shown in **Figure 5**.

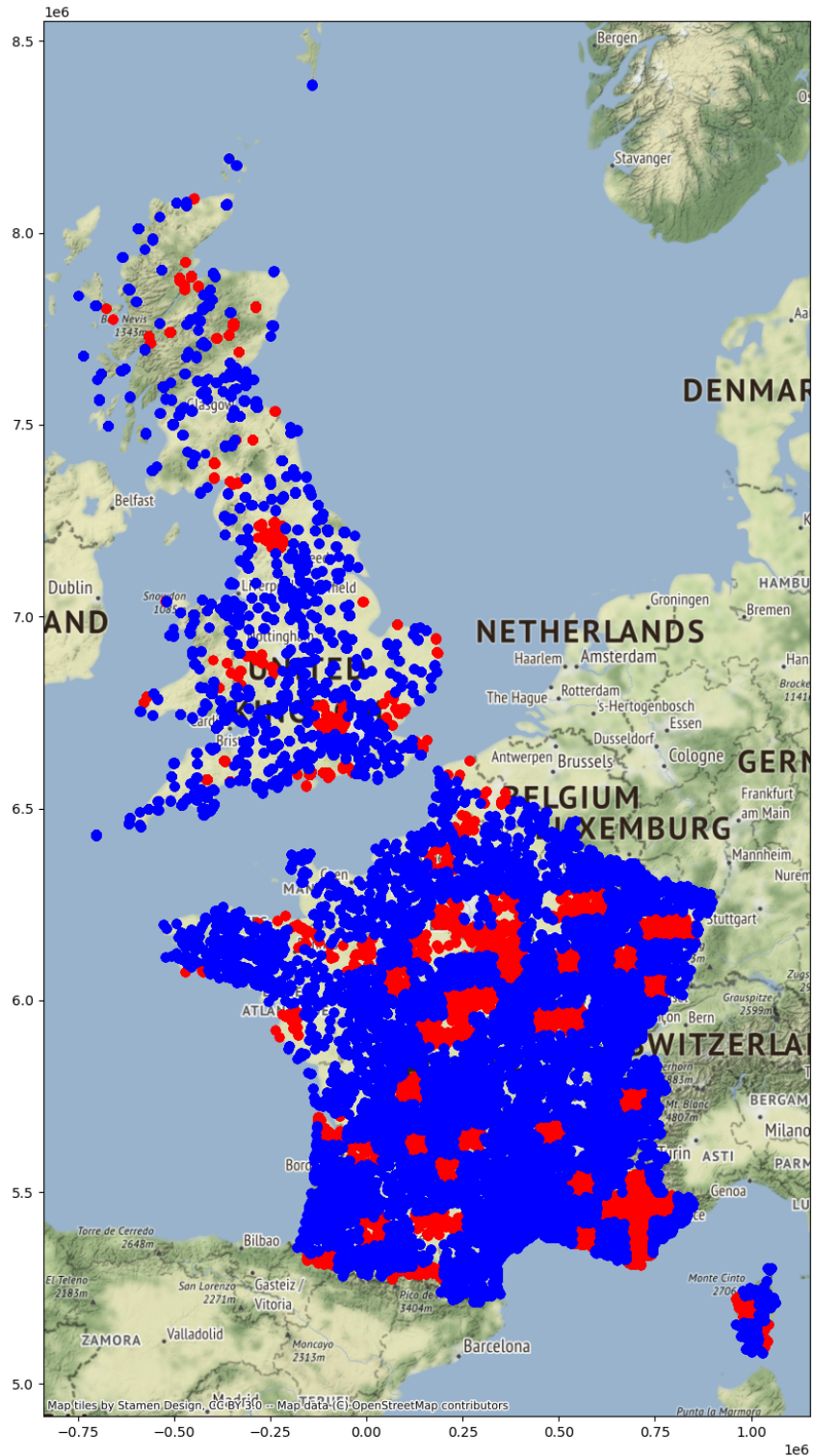

**Figure 5.** Locations of the 5.9 thousand validation (red) and 22.4 thousand test (blue) presence-absence (PA) surveys covering France and Great Britain.

4. Evaluation

The species identifiers associated with the test PA surveys were kept secret for the fair evaluation process of the GeoLifeCLEF campaign (Botella et al., 2023). However, anyone is welcome to freely evaluate its own predictions on the test set at any time through a late submission on the Kaggle page: https://www.kaggle.com/competitions/geolifeclef-2023-lifeclef-2023-x-fgvc10. The evaluation metric (F1 micro score) is explained and the results of many methods were analyzed in Botella et al. (2023).

5. Environmental predictor data

The spatialized geographic and environmental predictor data is to be used as input variables for the predictive approaches. For each species observation (PO or PA), we provide a four-band 128x128 satellite image at 10 m resolution around the occurrence location and time series of the past values for six satellite bands at the point location. Besides, we provide various environmental rasters at the European scale (e.g., climatic, soil, elevation, land use and human footprint variables) from which one may extract local 2D images and build 3D tensors using the data loader provided in the code (see **7 - Python data loaders**). These three types of input environmental data are illustrated in **Figure 6**. It is also possible to extract a monthly time-series of four climatic variables for any observation as we provide monthly climatic rasters from 2000 to 2019.

**Satellite imagery**:

- Sentinel-2 patches: 128x128 RGB and NIR images (hereafter patches) with a 10 m pixel resolution and a total of four bands, centered at the observation geolocation and taken the same year. Each species observation is associated with one RGB file with JPEG compression for the RGB patch and a grayscale one for the Near-Infrared. These patches were extracted from the Sentinel-2 rasters pre-processed by the Ecodatacube platform, accessed in January 2023, and are accessible at /SatelliteImages/.

- Landsat time series: Each observation is associated with the time series of the satellite median point values over each season since winter 1999 for six satellite bands (RGB, NIR, SWIR1 and 2). One CSV file was created for each band, where rows are the timeSerieIDs (identifying a combination of location and day, see subsection **v-**) and the columns are the 84 seasons from winter 2000 until autumn 2020. The original Landsat satellite data had a 30m spatial resolution. It was pre-processed by the Ecodatacube platform from which we extracted the time series in February 2023. The time series CSV files are accessible at /SatelliteTimeSeries/.

**Monthly climatic rasters** of four climatic variables (mean, minimum and maximum temperature and total precipitation) from January 2000 to December 2019, yielding 960 rasters covering Europe under the WGS84 coordinate system with a pixel resolution of 30 arcsec (~1 km). The rasters are GeoTIFF files with compression. They were extracted from Chelsa Climate in December 2022 and are accessible at /EnvironmentalRasters/Climate/Climatic_Monthly_2000-2019.

**Environmental rasters** of many variables, all provided in longitude/latitude coordinates (WGS84) with the same spatial extent including all the species observation data. All the variables are listed in **Table 2**.

- Bioclimatic rasters: 19 low-resolution rasters covering Europe. These 19 variables are classically used in species distribution modeling. They are provided as GeoTIFF files with compression and a resolution of 30 arcsec (~ 1 km). They were extracted from Chelsa Climate in December 2022 and are accessible at /EnvironmentalRasters/Climate/BioClimatic_Average_1981-2010.
- Soil rasters: Nine pedologic low-resolution rasters covering Europe. These variables describe the soil properties from 5 to 15 cm depth and are determinants of plant species distributions. They are provided as GeoTIFF files with compression and 30 arcsec resolution (~ 1 km). They were extracted from Soilgrids in December 2022 and are accessible at /EnvironmentalRasters/Soilgrids.
- Elevation: High-resolution raster of elevation (in m) covering Europe. It is provided as one GeoTIFF file with compression and Int16 numeric storage (13.2 GB). Resolution: 1 arcsec (~30 m). It was extracted from ASTER Global Digital Elevation Model V3 in January 2023 and is accessible at EnvironmentalRasters/Elevation.
- Land Cover: A medium-resolution multi-band land cover raster covering Europe. Each band describes either the land cover class prediction or its confidence under various classifications. We recommend the use of IGBP (17 classes) or LCCS (43 classes) layers, often used in species distribution modeling. It is provided as a GeoTIFF file with compression and a resolution of ~500m. It was extracted from MODIS Terra+Aqua 500m in December 2022 and is accessible at /EnvironmentalRasters/LandCover/.
- Human footprint: 16 low-resolution rasters describing human footprint, including 14 rasters of detailed pressures, i.e. seven pressures on the environment (e.g., nightlight level, population density) induced by human presence and activity times two time periods (~1993 and ~2009), and two rasters summarizing all pressures for the two time periods. This temporal redundancy allows the assessment of spatial and temporal changes in human pressures (Venter et al. 2016b). They are provided as GeoTIFF files with compression with a 30 arcsec resolution (~1 km). They were obtained from Venter et al., 2016 in January 2023 and are accessible at /EnvironmentalRasters/HumanFootprint/ where the subfolder summarized/ contains the two summary rasters and and detailed/ contains the 14 single pressure rasters.

### A. Satellite image patches

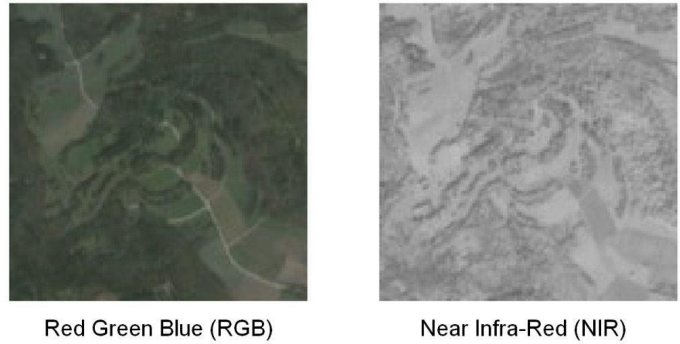

Red Green Blue (RGB)  Near Infra-Red (NIR)

### B. Satellite time series

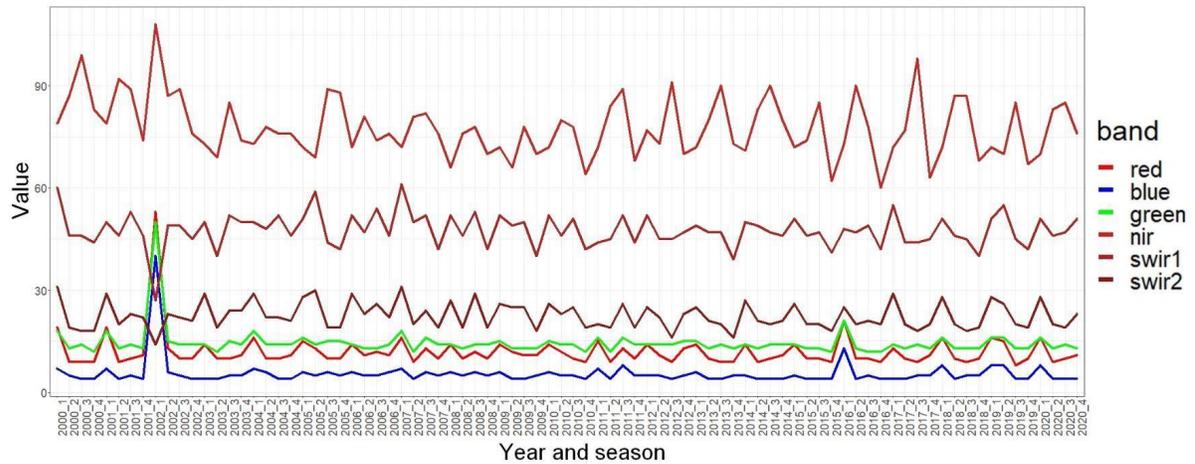

### C. Environmental rasters

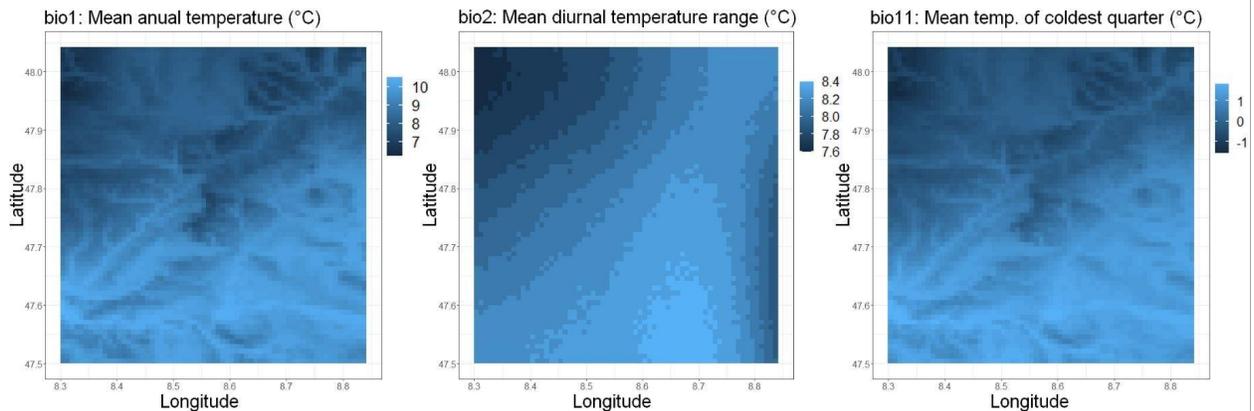

**Figure 6.** Illustration of the three format of geographic environmental data provided as input for the species distribution models: (A) a four band (RGB+NIR) satellite image (128x128) at ten meter resolution centered around each species observation geolocation, (B) a six band (RGB + NIR + SWIR1/2) monthly satellite time series covering 2000-2020 for the point location and (C) rasters (64x64 km) of three environmental (climatic) variables centered at the point location extracted from European scale rasters.

| Type | File name | Description | Source | spatial resolution |
|---|---|---|---|---|
| Bioclimatic average | bio1.tif | Mean annual air temperature | Chelsa Climate | 30 arc s. |
| | bio1.tif | mean annual air temperature | | |
| | bio2.tif | mean diurnal air temperature range | | |
| | bio3.tif | isothermality | | |
| | bio4.tif | temperature seasonality | | |
| | bio5.tif | mean daily maximum air temperature of the warmest month | | |
| | bio6.tif | mean daily minimum air temperature of the coldest month | | |
| | bio7.tif | annual range of air temperature | | |
| | bio8.tif | mean daily mean air temperatures of the wettest quarter | | |
| | bio9.tif | mean daily mean air temperatures of the driest quarter | | |
| | bio10.tif | mean daily mean air temperatures of the warmest quarter | | |
| | bio11.tif | mean daily mean air temperatures of the coldest | | |

| | | quarter | | |
|---|---|---|---|---|
| | bio12.tif | annual precipitation amount | | |
| | bio13.tif | precipitation amount of the wettest month | | |
| | bio14.tif | precipitation amount of the driest month | | |
| | bio15.tif | precipitation seasonality | | |
| | bio16.tif | mean monthly precipitation amount of the wettest quarter | | |
| | bio17.tif | mean monthly precipitation amount of the driest quarter | | |
| | bio18.tif | mean monthly precipitation amount of the warmest quarter | | |
| | bio19.tif | mean monthly precipitation amount of the coldest quarter | | |
| Bioclimatic monthly | XX_YYYY_pr.tif | Precipitation amount in kg m$^{-2}$ month$^{-1}$/100 | | |
| | XX_YYYY_tasmax.tif | Mean daily maximum 2m air temperature in K/10 | | |
| | XX_YYYY_tas.tif | Mean daily air temperature in K/10 | | |
| | XX_YYYY_tasmin.tif | Mean daily minimum air temperature in K/10 | | |
| Elevation | ASTER_Elevation.tif | Elevation | ASTER Global Digital | 1 arc s. |

|  |  |  | Elevation Model v003 |  |
|---|---|---|---|---|
| Human footprint | HFP1993_WGS84.tif | The Human Footprint camp of cumulative pressures on the environment in 1993 | Dryad | Data -- Global terrestrial Human Footprint maps for 1993 and 2009 | 30 arc s. |
|  | HFP2009_WGS84.tif | The Human Footprint camp of cumulative pressures on the environment in 2009 |  |  |
|  | Built1994_WGS84.tif | Individual pressure of built environments in 1994 |  |  |
|  | Built2009_WGS84.tif | Individual pressure of built environments in 2009 |  |  |
|  | Croplands1992_WGS84.tif | Individual pressure of croplands in 1992 |  |  |
|  | Croplands2005_WGS84.tif | Individual pressure of croplands in 2005 |  |  |
|  | Lights1994_WGS84.tif | Individual pressure of night-time lights in 1994 |  |  |
|  | Lights2009_WGS84.tif | Individual pressure of night-time lights in 2009 |  |  |
|  | Navwater1994_WGS84.tif | Individual pressure of navigable waterways in 1994 |  |  |
|  | Navwater2009_WGS84.tif | Individual pressure of navigable waterways in 2009 |  |  |
|  | Pasture1993_WGS84.tif | Individual pressure of pasture lands in 1993 |  |  |
|  | Pasture2009_WGS84.tif | Individual pressure of pasture lands in 2009 |  |  |

|  | Popdensity1990_WGS84.tif | Individual pressure of human population density in 1990 | | |
| --- | --- | --- | --- | --- |
|  | Popdensity2010_WGS84.tif | Individual pressure of human population density in 2010 | | |
|  | Railways_WGS84.tif | Individual pressure of railways circa 1990 | | |
|  | Roads_WGS84.tif | Individual pressure of roads circa 2000 | | |
| Land cover | LandCover_MODIS_Terra-Aqua_500m.tif | Multi-band raster file including the IGBP classification (17 land cover classes). | MODIS/Terra+Aqua Land Cover Type Yearly L3 Global 500m SIN Grid - LAADS DAAC | 500 m |
| Soil | crop_proj_bdod_5-15cm_mean_1000.tif | Bulk density (cg/cm3) | SoilGrids | 30 arc s. |
|  | crop_proj_cec_5-15cm_mean_1000.tif | Cation exchange capacity at ph 7 (mmol(c)/kg) | | |
|  | crop_proj_cfvo_5-15cm_mean_1000.tif | Coarse fragments in cm3/dm3 | | |
|  | crop_proj_clay_5-15cm_mean_1000.tif | Clay content in g/kg | | |
| | crop_proj_nitrogen_5-15cm_mean_1000.tif | Nitrogen in cg/kg | | |
| | crop_proj_phh2o_5-15cm_mean_1000.tif | pH water (pH *10) | | |
| | crop_proj_sand_5-15cm_mean_1000.tif | Sand in g/kg | | |
| | crop_proj_silt_5-15cm_mean_1000.tif | Silt in g/kg | | |
| | crop_proj_soc_5-15cm_mean_1000.tif | Soil organic carbon (dg/kg) | | |
| Sentinel-2 images | patchs_rgb.zip | Sentinel-2 RGB and NIR patches centered in the observation | EcoDataCube | 10 m |
| | patch_nir.zip | | | |
| Landsat time series | time_series_blue.csv | Landsat quarterly time series of six spectral bands (blue, green, red, NIR, SWIR1 and SWIR2), going back up to 20 years. | EcoDataCube | 30 m |
| | time_series_green.csv | | | |
| | time_series_red.csv | | | |
| | time_series_nir.csv | | | |
| | time_series_swir1.csv | | | |

|  | time_series_swir2.csv |  |  |  |

**Table 2.** The environmental variables available for each species observation in GeoLifeCLEF 2023, along with their source and spatial resolution.

6. Dataset structure, file entries

The two species observation data CSV files are the entry point to the diverse environmental predictors. The tabular file *Presence_Absence_train.csv* lists the PO records while the file *Presences_Absences_train.csv* lists the PA survey records as illustrated by **Figure 7**. 15 columns are shared between these two files. The basic ones are the glcID (unique row observation identifier distinct between PA and PO), speciesId (identifies uniquely each species which are anonymized), lon, lat (unprojected WGS84 coordinates), geoUncertaintyInM, year (year where the data was sampled), dayOfYear (day in the year where the data was sampled, from 1 to 366) and datasetName. Some other columns link towards the other dataset components (e.g. predictors):

- **patchID**: identifies the unique combinations of longitude, latitude and year. This identifier is used to locate and name the satellite patch files as explained below.

- **timeSerieID**: Identifies the unique combinations of longitude, latitude, year and dayOfYear. One can obtain the satellite time series of a given species observation through its timeSerieID for any band (Red, Green, Blue, NIR, Swir 1, Swir2), by looking at the corresponding row in the band-dedicated CSV file, as illustrated in **Figure 7**.

- **gbifID**: This identifier links to the original species observation on gbif.org, and exists only for the observations that were extracted from GBIF. For instance, the GBIF webpage of the observation with a gbifID of **3473384501** is accessible at: https://www.gbif.org/occurrence/3473384501

As mentioned above, the test PA data has a slightly different structure because the species identifiers are kept secret for a fair model evaluation in the campaign GeoLifeCLEF. Hence, a table *test_blind.csv* is provided with 22,404 rows, corresponding to the test surveys, and 12 columns including: **lon, lat, geoUncertaintyInM, year, dayOfYear, datasetName, patchID, timeSerieID**. The table also contains a specific column named **Id**, which uniquely identifies the sample unit (combination of patchID and dayOfYear) to compute the GeoLifeCLEf 2023 evaluation score (see Botella et al., 2023). One can freely and fairly evaluate its own models on our test dataset through the late submission system of the Kaggle page:

https://www.kaggle.com/competitions/geolifeclef-2023-lifeclef-2023-x-fgvc10/. It allows a free and fair overall performance assessment and its comparison with the methods submitted during the campaign (synthesized in Botella et al. 2023).

**Sentinel2 satellite patches**. There is one RGB patch and one NIR patch for each patchID of the observation data. The patches are compressed in two zip files (*patchs_rgb.zip*, *patchs_nir.zip*) accessible in folder /SatelliteImages/. To access them, one first needs to download and decompress these archives, generating the folders ./rgb and ./nir respectively. Each contains around 4 millions JPEG images organized in a folder tree of depth two. Each JPEG file is named with a "patchID". To recover the RGB (resp. NIR) patch of a species observation, the patchID column of the occurrence CSV is the key to the file path with the rule '.../CD/AB/XXXXABCD.jpeg' as illustrated in **Figure 7**. In the case where the patchID has less than 4 decimals, just add zeros before to get the folder path.

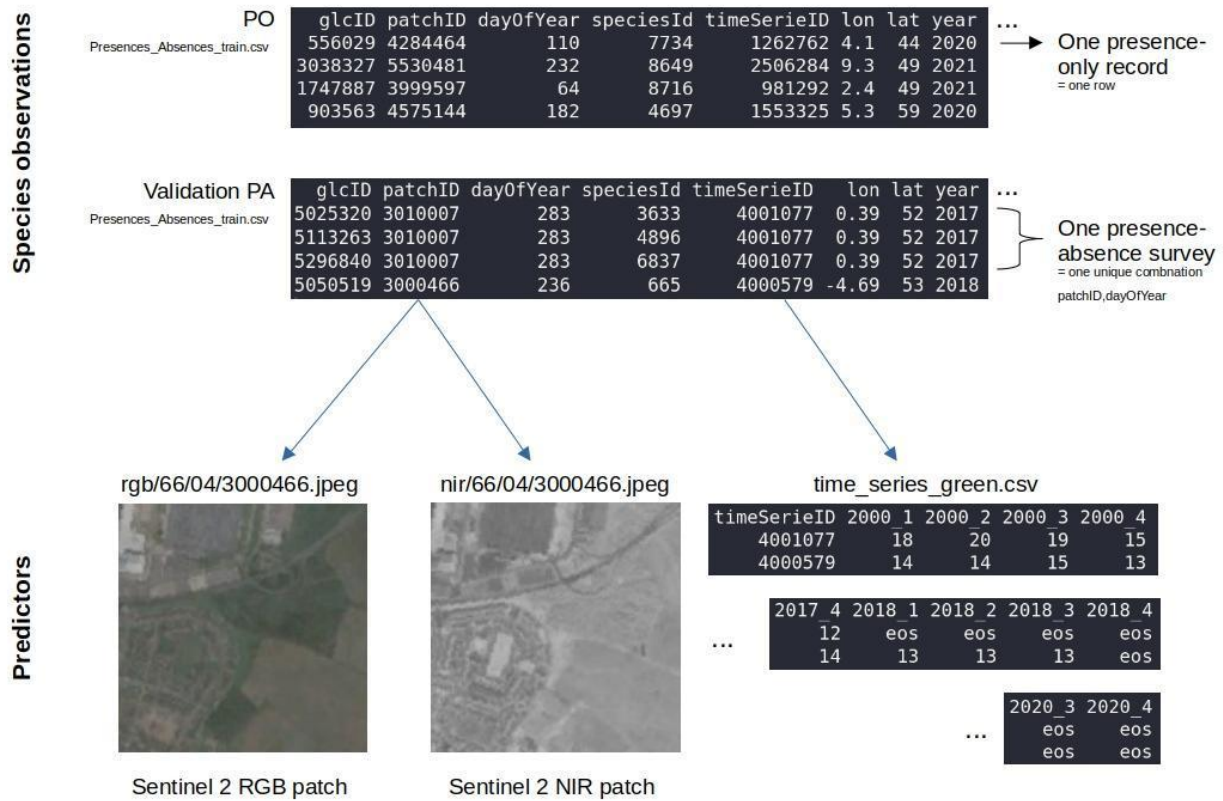

**Figure 7.** Illustration of species observation and satellite file names, headers and links in the GeoLifeCLEF 2023 dataset.

7. Python data loaders

So as to facilitate the use of the dataset, we developed tools to load the environmental predictors or species observations into data usable by deep learning frameworks (data loaders) and made available on the open GitHub repository (https://github.com/plantnet/GLC), along with sample data and plug-and-play examples (see **Figure 8**).

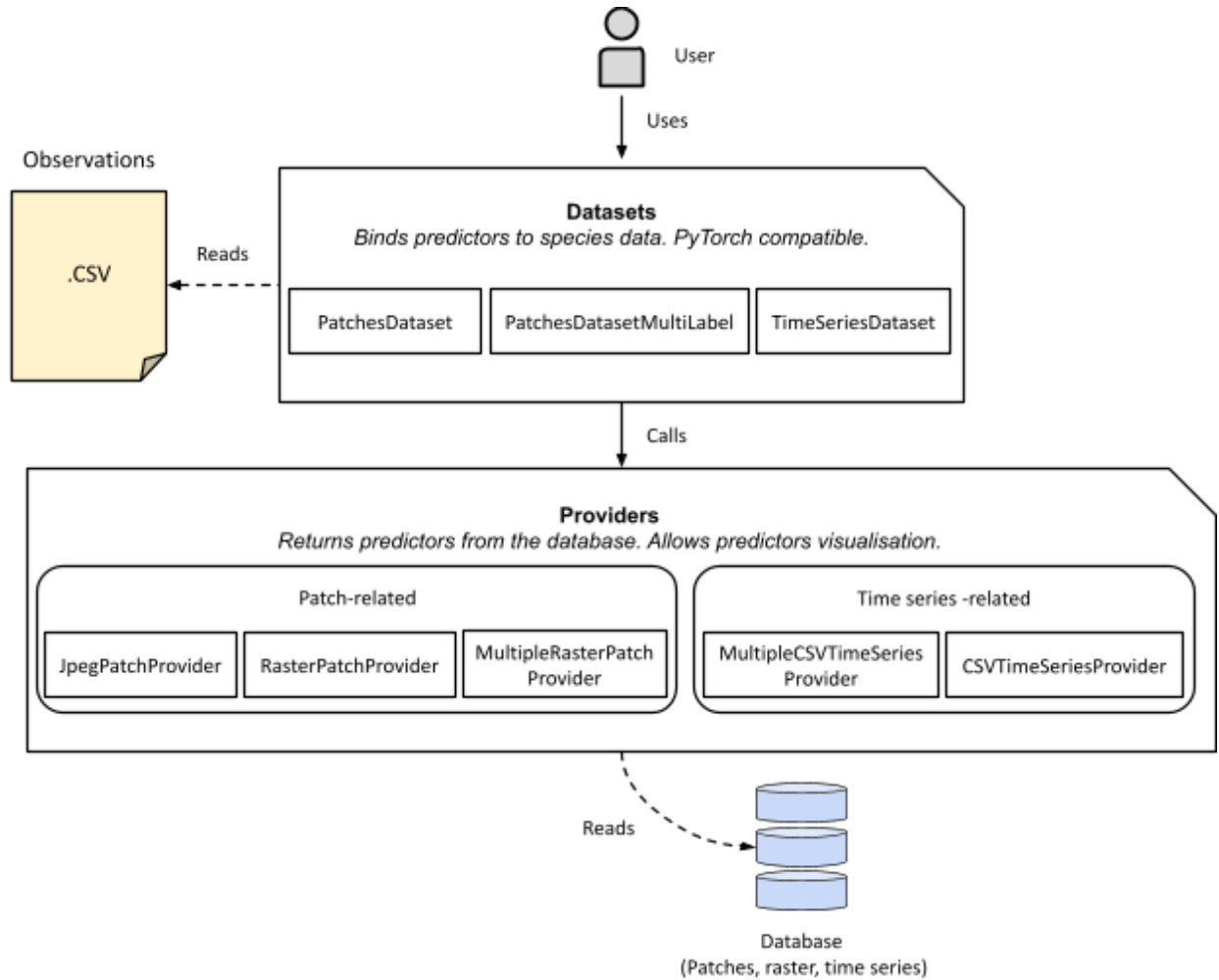

**Figure 8.** Main functionalities and components of the python data loader provided along with the dataset.

The data loaders have been developed to load either pre-extracted visual patches, visual / bioclimatic rasters or time series data; each coupled with an observation file containing geolocation information and predictor data identifiers (e.g. patchID / timeseriesID) tied to the desired data. By simply instantiating a Dataset object, with the desired sampler(s) and datapath(s), one can query the dataset with an integer index to retrieve an element (either a patch or a sequence) alongside its label (*i.e.* speciesId); or call the dataset's plot function with the same index to visualize the element in question. Furthermore, Dataset objects are PyTorch friendly and can be passed as input arguments of PyTorch functions for automated training and inference of deep neural networks.

# Methodology

1. Presence-only data

This data is composed of opportunistic geolocated observations of species whose sampling protocol is unknown and have a highly heterogeneous sampling coverage in space, time and across species. Indeed, these records are generally concentrated towards highly populated or easily accessible areas and tend to focus on charismatic and easy to identify plant species. Twelve PO datasets were pre-selected from GBIF (www.gbif.org) based on their amount of data, their geolocation accuracy and identification quality. To identify these datasets, we first searched gbif.org for occurrences with the following criteria: **Scientific name** belongs to *Tracheophyta*; 2021>=**year**>=2017; includes coordinates; geolocation uncertainty <100m; is in study area (cf 38 countries above). Among the returned datasets with the largest number of occurrences, we selected the ones that had a minimal documented validation procedure for species identification on the GBIF page or website (either automatic or by peers) and for which record's coordinates were not aggregated into grid cells nor blurred. Some datasets like wikiplantbase #Italia were excluded because no occurrence at all reported the geolocation uncertainty. We extracted the occurrences from GBIF.org (on 08 November 2022, https://doi.org/10.15468/dl.8wvzqf) with the criteria above and from these 12 selected PO datasets and two PA datasets (see Table 1), which covered together around 94% of the occurrences available with these criteria. Then, we sequentially applied the following pre-processing steps to the PO data:

1) We excluded occurrences whose taxon rank was higher than species (i.e. genus, family,etc).

2) Three datasets (totalling around 6 million occurrences) mixed different validation status and hence trust levels in their occurrences (Artportalen, Estonian and Waarnemingen.be). We used the field "Identification validation status" to exclude occurrences that were unvalidated, i.e. directly submitted by the original observer without any posterior validation.

3) We filtered duplicate observations by randomly keeping only one observation among the ones of the same species, observed the same day, by the same observer, in the same dataset and at the exact same location. Then, we also removed duplicates with the same species, location, day and observer between the "Pl@ntNet observations" and "Pl@ntNet

automatically identified" datasets because these duplicates were due to the posterior validation of an automatically identified record by the observer community.

4) We reduced the number of occurrences to 5 millions, by randomly selecting maxOcc occurrences for each species that had more than maxOcc occurrences after step (3). maxOcc is computed so that the final total number of occurrences equals 5 millions.

5) We computed the projected metric coordinates (EPSG:3035) in addition to the WGS84 longitudes and latitudes, and added it as columns x_epsg3035 and y_epsg3035 in the file presences_only_train.csv.

2. Presence-Absence data

The PA surveys were provided to control for the sampling biases in model calibration or through a joint statistical integration, as, for instance, using point processes (Fithian et al., 2015, Miller et al. 2019). Each PA survey corresponds to an exhaustive inventory of a small spatial plot (from 10m² to 400m²) by botanical experts. Spatial plots larger than 100m² are typically forest plots, for which it is common practice to increase the sampled area to better represent the local species assemblage. We included four source datasets, IGN, NPMS, CBNMed (extracted from SIMETHIS-Flore-CBNMed, Argagnon et al., 2022) and CBNA (**Table 1**) which sampled different areas and/or habitats. The IGN dataset has regular coverage across all the metropolitan French territory and specifically targets forest habitats, but sampling all the plants within a plot. The CBNmed and CBNA cover a smaller area, respectively the mediterranean French area (including Corsica) and the French Alps, but both sampled a diversity of habitats representative of their area with a relatively large species turnover between surveys. Finally the NPMS is a citizen science program that aims at covering all plant habitats across Great Britain with a standardized plot sampling protocol and its contributors received a prior botanical course. We only kept surveys whose geolocation uncertainty was inferior to 100m, and whose sampling year was between 2017 and 2021, consistently with the PO data. Besides, the CBNA and CBNMed data were originally not using the GBIF backbone taxonomy, so we matched their original taxa with the GBIF backbone taxonomy using the R package **rgbif**. If the matching was exact, we kept the matched species-level name, even when the original taxon was at higher level (subspecies, variety). If the matching was fuzzy, we compared the original name manually to its best match and kept the matched species-level name only if it appeared reasonable (45 original names were recovered this way). Otherwise, we removed the entries of the original taxon from the PA data. Then, we also filtered the species that were absent from the PO dataset. For the IGN, CBNA and CBNMed datasets, a spatial plot is mostly visited only once a year, while for the NPMS it is visited at two dates in nearly half of the cases.

3. Species anonymization

We have consistently anonymized species names after identifying them with integers (speciesId) in all species observations (PO and PA) on demand of certain PA data providers, in particular to avoid revealing precise locations of species of conservation concern while keeping all precise locations in the final dataset.

4. Split of validation and test data

A spatial block hold-out strategy was employed to limit the effect of the spatial auto-correlation in the data when evaluating the model (Roberts et al., 2017). To achieve this, the presence-absence surveys points are split according to a 50Km spatial grid (in EPSG:3035), with 80% of the grid cells (i.e. 22,404 surveys) assigned to the test set and the remaining 20% (i.e. 5,948 surveys) to the validation set (see **Figure 5**). In addition, to avoid any spatial overlap between the training set (presence only) and the test set, a filter is applied to the presence only data to remove all points that are less than 250 meters from a test set survey point.

5. Satellite data

Remote sensing data represent an unprecedented opportunity to provide high spatial resolution species distribution models with rich and globally-consistent predictor variables describing the environment (He et al, 2015).

- **Satellite image patches.** The Sentinel-2 satellite images span from 2017 to 2021 and are split between RGB and NIR bands. Ecodatacube pre-processed the data to eliminate cloud coverage and cloud shadow on tiles which were then merged into seamless mosaics, thus producing annual composite rasters. This composite is derived from the aggregation of the images of the year and is available in three versions: 25, 50 and 75 quantiles. In this dataset, only the 50 quantile (median) raster is used for each band. Sentinel data patches are extracted at each point with a coverage of 128x128 pixels, where pixel size is 10 x 10m. This patch is saved as a JPEG by changing the uint16 encoding (source encoding of EcoDataCube data) to uint8, to make the files lighter. To do this, the values are first thresholded at 10000, then rescaled to [0,1] and a gamma correction of 2.5 is applied (i.e. values are powered by 1/2.5). Finally the values are rescaled to [0,255] and rounded so that the final integer values allow the uint8 encoding. This process also avoids using range for high reflectance values (values >10000 in uint16) and gives more range to values close to zero which are the most common.
- **Satellite time series** The Landsat data span from 2000 to 2020, and were obtained from quarterly median composites provided by EcoDataCube, resulting in four values per year per band. By selecting the median (percentile 50) values along each three month period the chances of taking a cloud-free value are increased. A time series id was created to uniquely identify observations of a given geolocation and time window (quarters of each year). We then extracted the quarterly values for each location and filled the time series time steps corresponding to dates posterior to the observation with a special 'no data' token. This data carries a highly spatially localized signature of the past 20 years of seasonal vegetation changes, potential extreme natural events, such as fires, or land use changes.

6. Environmental rasters

All the source environmental rasters described below were re-projected, when necessary, to the WGS84 coordinate system (EPSG:4326), then cropped to the same spatial extent (file GLC23.geojson in our code repository, hereafter the GeoLifeCLEF bounding box) and saved as .TIF files. The bounding box was defined as the rectangle whose top-right corner would have a y-coordinate equal to the latitude of the northernmost species observation + 1 and a x-coordinate equal to the longitude of the easternmost observation + 1 and whose bottom-left corner would have a y-coordinate equal to the latitude of the southernmost observation - 1 and a x-coordinate equal to the longitude of the westernmost observation - 1 (min_x, min_y, max_x, max_y = -32.26344, 26.63842, 35.58677, 72.18392).

Bioclimatic average rasters. It was shown during the previous edition of the challenge (see Lorieul et al., 2022) that the information on climatic conditions was essential for the spatial prediction of plant (and animal) species (see also Leblanc et al., 2022). We integrated the 19 average bioclimatic rasters from CHELSA (Karger et al., 2017). These 19 variables (see **Table 1**) are averaged over the period 1981-2010 and are classically used for species distribution modeling. They represent notably mean annual temperature, annual precipitation, seasonality and extreme or limiting environmental conditions. The 19 rasters were originally in the WGS84 coordinate system and were simply cropped to the GeoLifeCLEF bound box.

Bioclimatic monthly rasters. We provided monthly climatic rasters which may be used to the succession of climatic conditions of any location by reconstructing its climatic time series. Four climatic variables were provided as monthly rasters (see **Table 2**). The data was retrieved from CHELSA (Karger et al., 2017) and consisted of .TIF files showing the means of calendar months per grid cell with a 30 arcsec resolution in the WGS84 coordinate system. In order to work with recent data, we only kept rasters from January 2000 to December 2019 (except for the **pr** layer which was missing between July and December 2019 when the data was accessed), which resulted in 4*12*20-6=954 files. Each of these source rasters was cropped to the GeoLifeCLEF 2023 bounding box and saved as a TIF file with the DEFLATE compression algorithm (see Oswal et al., 2016). The code can be found at */data_preparation/monthly_rasters_preparation.py*.

Soilgrids rasters. Physico-chemical soil properties (e.g. Ph, granulometry) are crucial determinants of plant species ability to survive. SoilGrids (Poggio et al., 2021) is a system for global digital soil mapping that uses machine learning methods to map the spatial distribution of soil properties across the globe. SoilGrids prediction models are fitted at 250m resolution using over 230 000 soil profile observations from the WoSIS database and a series of environmental covariates. We integrated nine soil rasters corresponding to a depth of 5 to 15cm at a resolution of 30 arc sec, i.e. the aggregated version of SoilGrids 2.0 rasters derived by resampling at 1km the mean initial predictions at 250m for each property. These rasters were downloaded in january 2023 from isrig.org, already under the WGS84 coordinate system, and cropped to the GeoLifeCLEF bounding box.

Human footprint rasters. The global terrestrial human footprint rasters from Venter et al. (2016a) provide reference data to account for human settlement and activities in species distribution modeling. The data are derived from both remote sensing and bottom-up surveys, and measure direct and indirect human pressures at the kilometer scale. It includes eight variables: 1) built environment, 2) population density, 3) electrical infrastructure, 4) cropland, 5) pastureland, 6) roads, 7) railways and 8) navigable waterways. With the exception of roads and railways, each variable is available and temporarily-consistent for both 1993 and 2009 (years are approximate, see Venter et al., 2016a). Synthetic human footprint rasters combining the eight variables are also provided by the authors for the two periods. The cumulative scores are normalized by biome, as described in Sanderson et al. (2002). This results in different absolute pressure levels in distinct biomes being equally represented and therefore needs to be acknowledged. The rasters were reprojected from Mollweide (ESRI:54009) to WGS84 geographic coordinate system (see script *data_preparation/reprojects_raster.py*) and then cropped to the GeoLifeCLEF 2023 bounding box as described above (*data_preparation/crops_HF_rasters.ipnyb*). Provided as deep SDM inputs, this information allows the description of human impact at the location of the observation. We provide two summary rasters combining all human pressures and two detailed rasters per pressure which avoid an arbitrary degradation of the original data. The human impact on biodiversity loss has been widely studied and recognised in international reports, with 1 million species already at risk of extinction (IPBES, 2019). Indeed, human pressure has been shown to drive changes in species extinction risk (Di Marco et al., 2018) and to be a better predictor of species' geographic range than biological traits (Di Marco et al., 2014). The inclusion of such a covariate in the GeoLifeCLEF 2023 dataset therefore seems essential.

Elevation raster. The Terra Advanced Spaceborne Thermal Emission and Reflection Radiometer (ASTER) Global Digital Elevation Model (GDEM) Version 3 (ASTGTM) provides a global digital elevation model (DEM) of land areas on Earth at a spatial resolution of 1 arc second (approximately 30 meter resolution at the equator). We extracted the tiles overlapping with the GeoLifeCLEF bounding box (shapefile *data_preparation/GLC23.geojson*) through the NASA earthdata portal (https://search.earthdata.nasa.gov). The TIF tiles were merged and converted to a single compressed Cloud Optimized .GeoTIFF (*data_preparation/merges_ASTER_elevation_tifs.py*). Topography has a distal effect on plant distribution as it influences light, moisture and nutrient conditions, among others. As expected, the inclusion of topography as a covariate in SDM has been shown to significantly improve its performance (Sormunen et al. 2011). Data on edaphic factors are still extremely scarce on a large scale and topography appears to be an excellent proxy for them.

Land cover raster. A stack of land cover rasters is provided at approximately 500m resolution for the whole study area and period 2017 to 2019 and originates from MODIS Terra+Aqua (Friedl et al., 2010). To obtain it, we extracted the 24 HDF raster tiles overlapping the GeoLifeCLEF bounding box from the NASA earthdata portal (https://search.earthdata.nasa.gov/search/, accessed on the 12th of January). These HDF files stack several 13 layers corresponding to various land cover classifications or encoding the class confidence detailed at: https://lpdaac.usgs.gov/documents/101/MCD12_User_Guide_V6.pdf. The HDF files were merged into a single multi-band GeoTIFF, which was reprojected to WGS84 and cropped using the GeoLifeCLEF bounding box (*data_preparation/merges_LandCover_hdfs.py*). Land cover variables help to explain species distributions at all scales and were shown to significantly improve bioclimatic model performance at thinner spatial resolutions starting from 20 km (Luoto et al., 2005). Furthermore, the interactions between climate change and land cover change remain poorly understood and could strongly modify both land cover change itself and the distribution of threats (Mantyka-Pringle et al., 2015).

## Data accessibility

The whole dataset and code is available on open repositories, except the species labels of the evaluation dataset, for research purposes. All the source datasets are coherent with this policy. Species are anonymized to prevent threats to sensitive species.

The data may be accessed at the following permanent link: https://lab.plantnet.org/seafile/d/936fe4298a5a4f4c8dbd/

The code for the data preparation and the data loaders is available at: https://github.com/plantnet/GLC

# Funding statement

The research described in this paper was funded by the European Commission via the GUARDEN (https://cordis.europa.eu/project/id/101060693) and MAMBO (https://doi.org/10.3030/101060639) projects, which have received funding from the European Union's Horizon Europe research and innovation programme under grant agreements 101060693 and 101060639.

# Literature Citation